\documentclass[aps,prb,preprint,floatfix]{revtex4}
\usepackage{graphicx}
\usepackage{dcolumn}
\usepackage{bm}
\begin{document}

\title{The constant-pressure molecular dynamics for finite systems and its applications}

\author{D. Y. Sun$^{1,2}$ and X. G. Gong${^1}$}
\address{${^1}$Department of Physics, Fudan University, Shanghai-200062, China}
\address{${^2}$ Key Laboratory of Optical and Magnetic Resonance
Spectroscopy and Department of Physics, East China Normal
University, Shanghai 200062, China}

\date{\today}

\begin{abstract}

Recently Sun and Gong proposed a new constant-pressure molecular
dynamics method for finite systems. In this paper, we discuss the
current understanding of this method and its technique details. We
also review the recent theoretical advances of nano-system under
pressure by using this method.

\end{abstract}

\maketitle

\begin{center}
{\bf $I$. Introduction}
\end{center}

Nowadays, the molecular-dynamics (MD) simulation, widely used in
chemistry, physics, and materials sciences, is considered as a
standard and powerful tool for investigating the structures and
properties of matters in atomic scale.\cite{AlT} As an important
improvement, the constant-pressure MD(CPMD) proposed by
Andersen,\cite{An} and subsequently extended by Parrinello and
Rahman,\cite{PaR} has opened a crucial window to explore systems
under the pressure and tensions. Over the past decades, CPMD plays a
key role for our understanding of many phenomena relevant to high
pressure experiments in atomic scales.

Although the traditional CPMD has archived great success for bulk
matters at high-pressure conditions, it fails to be directly used for
finite size systems and for systems without regular shapes, such as
nanocrystals, where the boundary is hard to describe. Motivated
by the experimental work for the molecular, low-dimensional,
biological system\cite{AnP,Ga} and nanocrsystals under
pressure,\cite{ToA} three theoretical groups proposed the CPMD for
finite systems. To achieve the goal, Martonak, Molteni and
Parrinello have made the first step by directly extending the
traditional CPMD.\cite{MaM} Hereafter we call this method as the
directed method. This method directly mimics a real high-pressure
experiment, \emph{i.e.} the system keeping at constant pressure
through exchange of linear momentum with environments (in their
paper the environment is the pressure transmitting liquid).
According to Martonak, Molteni and Parrinello,\cite{MaM} a target
cluster is immersed into a well-chosen pressure-transmitting liquid,
the whole system (liquid+cluster) is simulated using
Parrinello-Rahman CPMD. The Parrinello-Rahman CPMD is a well
developed technique, so the key point of using this method is the
choice of pressure-transmitting liquid, specifically the interaction
between liquid-liquid and liquid-cluster. In the real application,
these interactions should be set to prevent the liquid from being
inside the cluster, and from phase transition happening during the
simulation. Furthermore one should pay more attentions to the choice
of pressure-transmitting liquid when the target cluster being in
liquid state due to the diffusivity. In their paper, Martonak,
Molteni and Parrinello have used a classical repulsive liquid, and
with classical interactions between the cluster and the
liquid.\cite{MaM} The main drawback of this method is the number of
the pressure-transmitting liquid atoms should much larger than that
of the cluster. Additionally, the direct method is suffered the same
problem as the original CPMD by the artificial \emph{mass}
associated with the piston.

Kohanoff, Caro and Finnis presented another method by introducing
the stochastic Brownian forces to each surface atoms (hereafter
stochastic method).\cite{KohanoffCF} This method can be considered
as simplified version of the direct method. In this treatment, the
surrounding fluid was replaced by random forces, which only act on
the surface of clusters. This situation is equivalent to a Brownian
motion described by the stochastic Langevin equation, where random
forces replace the collisions with the fluid, and a constant viscous
force represents the drag of the cluster motion immersed in the
fluid. These two types of forces are related by the
fluctuation-dissipation theorem. Physically, the interaction between
the clusters and surrounding liquid is not fully stochastic, the
random forces using in this method should be carefully.

For the same purpose, Gong and Sun proposed an alternative CPMD for
finite system.\cite{Sun02} In their approach, the system Lagrangian
is extended to include the \emph{PV} term, where \emph{P} is the
external pressure and \emph{V} is the volume of nanoclusters. By
writing the volume as a function of atomic coordinates, the constant
pressure can be readily achieved without any pressure-transmitting
liquid and without any artificial parameters. Hereafter this method
is named as extended method. In the application level, the key issue
of this method is to express the cluster volume as a proper function
of atomic positions. Since without periodic boundary conditions,
even without a regular shape, it is nontrivial to get a proper
definition of volumes. Gong and his co-workers have proposed a few
definitions. In the original CPMD paper,\cite{Sun02} Gong \emph{et al}
decomposed the cluster volume as the summation of individual atoms,
which has been used for metallic systems. Another definition due to
Gong and his co-worker is the cluster approximated by an ellipse, thus the
volume can be written in term of the principal radii of gyration. To
calculate the enthalpy of clusters, Calvo and Doye\cite{Calvo04}
give a more precise definition of volume as the minimum polyhedron
enclosing the cluster.

Carefully using these new methods, now it is possible to make
theoretical calculation for the low-dimensional system under
pressure.\cite{ToA} Over the past years, the extended method has
been used to study the structure and elastic properties of silicon
clusters,\cite{Ji_un} metallic clusters,\cite{Ye_un}
nanotubes,\cite{Sun04,Ye05,Ye07a} CdSe nanocrystals,\cite{Ye07b}
$C_{60}$ and diamond clusters,\cite{Baltazar06} \emph{etc}. The
directed method has been successfully used for the nano systems
including the silicon clusters,\cite{MaM} nanotube,\cite{Tangney05}
CdSe nanocrystal,\cite{Grunwald06}  $C_{60}$ and diamond
cluster,\cite{Baltazar06} \emph{etc}. The Stochastic method has been
employed to the Au clusters.\cite{KohanoffCF}

In this paper, we have reviewed the current understanding about the extended
method, as well as several recent theoretical advances of nano
systems under pressures. The technique details of the extend CPMD
(ECPMD) are presented in section $II$;  The numerical tests of ECPMD
are presented in section $III$; In section $IV$, the definition of
the volume for nano systems are recalled; In section $V$, the
application of the new method for some nanocrystals and nanotubes
are presented; Finally, we summarize the major conclusions in
section $VI$.

\begin{center}
{\bf $II$. The extended constant-pressure MD(ECPMD) method}
\end{center}

Considering a real $N$-atom  system, its Lagrangian $L_{real}$
takes:
\begin{equation}
L_{real}=\sum_{i}^{N}\frac{{\bf p}_{i}^{2}}{2m_{i}}-\phi(\{{\bf
r_{i}}\})
\end{equation}
where $\bf{r}_{i}$, $m_{i}$, ${\bf p}_{i}$ are the position, mass,
momentum of $ith$ atom respectively, and $\phi$ is the interaction
potential. In the extended method introduced by
Gong and Sun,\cite{Sun02} the system is extended to include a $PV$ term, and
Lagrangian $L_{extend}$ reads:
\begin{equation}
L_{extend}=\sum_{i}^{N}\frac{{\bf p}_{i}^{2}}{2m_{i}}-(\phi(\{{\bf
r_{i}}\})+P_{ext}V)
\end{equation}
where $V$ and $P_{ext}$ are the volume of the system and the
external pressure respectively.

The equations of motion(EOM) for the extended system derived from the
Lagrangian $L_{extend}$ are,

\begin{equation}
\frac{d}{dt}(\frac{\partial L_{extend}}{\partial {\bf {\dot
r}}_{i}}) = \frac{\partial L_{extend}}{\partial {\bf r}_{i}}
\end{equation}

The forces acting on the atoms compose of two parts, \emph{i.e.},
the force due to interatomic potential ($\overrightarrow{f}_{I}$)
and the one due to the PV term ($\overrightarrow{f}_{PV}$). EOM
derived from Eq. 3 produces the constant pressure ensemble for the
real systems, which can be obtained according to virial theorem.
\begin{equation}
<\frac{1}{3V}(\sum_{i}^{N}m_{i}v_{i}^{2}-\sum_{i}^{N}{\bf r}_{i}
\cdot {\bigtriangledown_{i}}\phi -\sum_{i}^{N}{\bf r}_{i}\cdot
P_{ext}{\bigtriangledown_{i}}V)>=0
\end{equation}
where $v_{i}$ is the velocity of $ith$ atom, and $<>$ denotes
average. Then we have,
\begin{equation}
<\sum_{i}^{N}m_{i}v_{i}^{2}-\sum_{i}^{N}{\bf
r}_{i}\cdot{\bigtriangledown_{i}}\phi>= <\sum_{i}^{N}{\bf
r}_{i}\cdot P_{ext}{\bigtriangledown_{i}}V>
\end{equation}

In the classic statistical physics, there is a basic assumption that
any statistical result can be obtained exactly from Newton's
mechanics, and any statistical quantity should be a function of
coordinates and velocity of atoms.\cite{LaL} Obviously the volume
can be written as a cubic homogeneous function of atomic positions.

\begin{equation}
V=V(......,r_{i}^{3},.....),
\end{equation}
where $r_{i}$ is the position of the $i$th atom. In fact, all the
occupied space by a cluster  can be divided up into tetrahedra with
atoms at their corners, thus the volume of a cluster is the cubic
homogeneous function of three Cartesian components of atomic
positions. Let $(x_{i},y_{i},z_{i})(i=0\sim3)$ denote the Cartesian
coordinates of the four vertexes of a tetrahedron, thus the volume
can be written as,

\begin{equation}
V_{tetrahedron}=\frac{1}{6}\left|
               \begin{array}{ccc}
    x_{0}-x_{1} & y_{0}-y_{1} & z_{0}-z_{1} \\
    x_{0}-x_{2} & y_{0}-y_{2} & z_{0}-z_{2} \\
    x_{0}-x_{3} & y_{0}-y_{3} & z_{0}-z_{3} \\
               \end{array}
             \right|
\end{equation}
The total volume is the summation of the each individual
tetrahedron,$V=\sum V_{tetrahedron}$.

According to Euler theorem,
\begin{equation}
\sum_{i}^{N}{\bf r}_{i}\cdot{\bigtriangledown_{i}}V=3V
\end{equation}
Finally, we end up with
\begin{equation}
P_{ext}=P_{int}=<\frac{1}{3V}(\sum_{i}^{N}m_{i}v_{i}^{2}-\sum_{i}^{N}
{\bf r}_{i}\cdot{\bigtriangledown_{i}}\phi)>
\end{equation}
where $P_{int}$ refers to the internal pressure, since
the external pressure $P_{ext}$ is
a constant,
$P_{int}$ is also a constant.
Thus,  by writing the volume as a function of atomic coordinates,
 the constant-pressure MD is achieved.

In some special cases, the
system size in certain directions fixed( in other words, the volume
of systems is independent of the atomic position along the special
direction). For this system, external pressure corresponds to an uniaxial pressure.
The uniaxial pressure can be realized by only including the one or two components of
$\overrightarrow{f}_{PV}$. For example, If only x-component of
$\overrightarrow{f}_{PV}$ includes in the simulation, this means the volume is
independent on the y and z components. Now
equation 5 becomes
\begin{equation}
<\sum_{i}^{N}m_{i}v_{xi}^{2}-\sum_{i}^{N}{\bf
x}_{i}\cdot\bigtriangledown_{xi}\phi>= <\sum_{i}^{N}x_{i}
P_{ext}\frac{\partial V}{\partial x_{i}}>
\end{equation}
Then we have,
\begin{equation}
P_{ext}=P_{xx}=<\frac{1}{V}(\sum_{i}^{N}m_{i}v_{xi}^{2}-\sum_{i}^{N}
{\bf r}_{i}\cdot{\bigtriangledown_{xi}}\phi)>
\end{equation}
where $P_{xx}$ is pressure along x-direction.

Similarly if
only x and y-component of $\overrightarrow{f}_{PV}$ includes in the
simulation, this means the volume is
independent on the z component.  Now we have,
\begin{equation}
P_{ext}=\frac{1}{2}(P_{xx}+P_{yy})=<\frac{1}{2V}(\sum_{i}^{N}m_{i}(v_{xi}^{2}+v_{yi}^{2})-\sum_{i}^{N}
{\bf x}_{i}\cdot{\bigtriangledown_{xi}}\phi+{\bf y}_{i}\cdot{\bigtriangledown_{yi}}\phi)>
\end{equation}
The above equation has been used in the study of nanotubes under
radial pressures.\cite{Sun04,Ye05,Ye07a} It also can be used for the
surface systems.

Combining with the constant temperature method, the constant
pressure method could be readily extended to constant-temperature
and constant-pressure ensemble. This extension is straight forward.
Most simulations at finite temperature in this paper are preformed
by combining with a Nos$\acute{e}$-Hoover thermostats.\cite{NH} The
extension of the method to $ab-initio$ molecular dynamics is also
simple.\cite{Ji04}

The ECPMD method is different from Andersen-Parrinello-Rahman
CPMD(APR-CPMD) physically, which has been misunderstood by a few
authors. First all, the volumes in ECPMD and APR-CPMD play the different role. In
APR-CPMD method, the volume is a generalized coordinates, which has
equal importance as an atomic coordinate.
However in ECPMD scheme, the volume is just a function of atomic
coordinations, even not a dynamics variable.
Secondly, the Lagrangian includes a virtual kinetic energy and mass associated
with the volume in APR-CPMD, which is absented in ECPMD, thus the atomic
dynamics in the methods could be different.
Finally, in APR-CPMD, the responding of system to the external pressure is
essentially linear and global, $i.e.$, all the atomic position is
linearly scaled in the same time. However, in ECPMD, the responding
of systems to the external pressure is truly local and non-linear.
This is especially important for the inhomogeneous system.
It needs to point that, although the two approaches could be
different in dynamics level, in the thermodynamics level, both do
realize the constant pressure ensemble.

\begin{center}
{\bf $III$. The volume of a cluster.}
\end{center}

One of key issues of the ECPMD is to properly define the volume for
a finite system. To do this, one should keep two points in mind. One
is the intrinsic uncertainty due to thickness of cluster surface, which
depends on what kinds of materials used to explore its thickness.
The previous studies on the cluster\cite{cheng92}, nanotube and nanowall\cite{Lu97}
have met this
problem. Another one comes from the computational consideration.
Geometrically, one can calculate the volume for any cluster, but it
is non-trivial to find one easy to implement and computational
cheap.

Before discussing the specific definition of volumes, we would like
to make some general comments. First, the force due to the \emph{PV}
term just acts on the surface atoms, because only the motion of
surface atoms directly changes the volume of systems. This is
consistent with the fact that the pressure-transmitting liquid only
interacts with the surface atoms. Secondly, the calculation
of volume could be much different specific forms, however as only as
each different form gives the same volume for all the
configurations, these forms will produce the same dynamics.
This is easy to understand mathematically.

More generally, let $V_{1}$ and $V_{2}$ to be two different
definitions, and $V_{2}$=$aV_{1}+b$, where \emph{a} and \emph{b} are
constants. The partition function calculated by $V_{1}$ and $V_{2}$ has following relationship,
\[Z_{2}(P)=\int e^{-\beta (\phi+PV_{2})}\prod dr_{i}=\int e^{-\beta (\phi+P(aV_{1}+b))}\prod dr_{i}\]
\[=\int e^{-\beta (\phi+aP(V_{1}+b/a))}\prod dr_{i}=e^{-\beta Pb/a}\int e^{-\beta (\phi+aPV_{1})}\prod dr_{i}=e^{-\beta Pb/a}Z_{1}(aP) \]
Supposing $A_{1}$ and $A_{2}$  are the ensemble average of a
physical quantity obtained by using $V_{1}$ and $V_{2}$
respectively. According to statistical physics, the thermal average
of a physics quantity A reads,
\begin{equation}
A_{1}(P)=\frac{1}{Z_{1}}\int A(\{r_{i}\})e^{-\beta
(\phi+PV_{1})}\prod dr_{i},
\end{equation}
\begin{equation}
A_{2}(P)=\frac{1}{Z_{2}}\int A(\{r_{i}\})e^{-\beta
(\phi+PV_{2})}\prod dr_{i} =\frac{1}{Z_{1}}\int
A(\{r_{i}\})e^{-\beta (\phi+aPV_{1})}\prod dr_{i}
\end{equation}
where $\beta$=1/$k_{B}T$, $Z_{1}$ and $Z_{2}$ are the partition
function corresponding to $V_{1}$ and $V_{2}$ respectively.
Comparing above two equations, one can easily conclude that
\begin{equation}
A_{2}(P)=A_{1}(aP).
\end{equation}
It implies the physics could be the same for the two
different definition of volumes, but it may happen in different
pressures, if the two definitions have the linear relationship.
Eq.15 also provides a very useful tool for comparison MD
results, where different volumes are used.

If the volume uncertainty due to surface could be neglected, the
exact volume can be calculated in principle. One of the very
accurate definition of volume is writing the volume as the summation
of all the no cross tetrahedron formed by four atoms,

\begin{equation}
V=\sum V_{tetrahedron},
\end{equation}
where $V_{tetrahedron}$ can be calculated using Eq.7. Although, this
scheme for volumes is exact, in computational view, it brings large
overloading for simulations. In fact, this method was not found
in any real simulations.

From computational viewpoints, one usually needs to find a more
reliable and cheap way to calculate volumes. One of
the simple and sufficient ways is to approximate the volume of each
atom based on the Wigner-Seitz sphere, \emph{i.e}, the scaled volume
of the atomic sphere to replace the Wigner-Seitz primitive
cell(hereafter labeled as \emph{VWS}), which has the following form,
\begin{equation}
V_{i}= \gamma_{i} \frac{4\pi}{3N_{i}}\sum_{j\neq
i}(\frac{r_{ij}}{2})^{3}, r_{ij}<r_{c}
\end{equation}
here $r_{c}$ keeps between 1st and 2nd nearest neighbors, $N_{i}$ is
the numbers of the nearest neighbor of the $i$th atom, and the
summation runs over all the first nearest neighbors of the $i$th
atom, $\gamma_{i}$ is a scale factor. For close packed structure,
$\gamma_{i}$ is approximated to have the value of 1.353.

\emph{VWS} was found to work well for metals.
Fig.1 (middle panel) shows the exact volume and one calculated by
\emph{VWS} for bulk Ni liquid at 3000K and 5GPa. One can see that the
volume calculated by \emph{VWS} does not recover the exact one
instantly. However the instant fluctuation can be much reduced by a
short time average. We find that the short time average is in
excellent agreement with the exact one (up panel of Fig.1). Since
most physical quantities are calculated through time average, we
believe the instant fluctuation will result in little effect on
physical results.

For most clusters, the ellipsoid is a good approximation to its shapes,
its volume can be also approximated by the volume of the
ellipsoid. The volume of an ellipsoid is determined by three
semi-axes, which can be given by the radii of gyration $R_{i}
(i=1,2,3)$ of this cluster. This definition of volumes was first
used in studying the glass transition of $Al_{n}$ clusters by Sun
and Gong,\cite{Sun98} and recently extended by Baltazar \emph{et
al.}.\cite{Baltazar06} According this definition, the volume of the cluster is,
\begin{equation}
V= C\frac{4\pi}{3}R_{1}R_{2}R_{3}
\end{equation}
Where $C$ is a scaling constant, which can be adjusted appropriately
according to its real volume. Following Baltazar \emph{et
al.},\cite{Baltazar06} the volume can be re-expressed as,
\begin{equation}
V= C\frac{4\pi}{3}\sqrt{\frac{det(I)}{N^{3}}}
\end{equation}
where det(\emph{I}) is the determinant of the inertia tensor
\emph{I}. This definition was found to work quite well for $C_{60}$ and Si nanocrystal.\cite{Baltazar06}
It is also recommended for metal systems.

As we mentioned above, the volume is only determined by the position
of surface atoms, thus the volume can be written as the minimal
polyhedron enclosing the finite system. This definition have been
used by Calvo and Doye to calculate the cluster enthalpy.\cite{Calvo04}
In using this method, one
should pay special attention for clusters with large negative surface
curvature, since it is easy to judge the surface atoms
for a cluster with positive structural curvature, but it may be much
subtle for part of clusters with negative structural curvature. For
finding the minimal polyhedron, the most used one is called the
quick convex hull algorithm.\cite{Barber96} Recently, this approach
has been used for studying the structure transition of CdSe
nanocrystal.\cite{Ye_un}

For some special structures, a specific
definition of volume will much simply the computing.
For example, the surface atoms of nanotubes and fullences can be easily located,
the volume of a nanotube can be defined
through the minimal polyhedron method.

\begin{center}
{\bf $IV$. The numerical tests of ECPMD}
\end{center}

The equation 3 does produce the constant pressure ensemble as shown
below. As an example, we simulated the carbon nanotube(CNT) and
$C_{60}$ at 300K based on equation 3. In this study, the volume is
defined through the minimal polyhedron method, and the temperature
is maintained by Nos$\acute{e}$-Hoover thermostate.\cite{NH} The
interaction between carbon atoms is described by a parameterized
many-body potential.\cite {Te,Br} In the calculation for CNT, the
pressure is only applied to all directions normal to axis. In Fig.2,
we present the volume, enthalpy and pressure as a function of times
for $C_{60}$. From this figure, we can see that the evolution of the
instantaneous volume, pressure and enthalpy fluctuates around the
average value, and the average pressure equals to the applied
external pressure. The correlation between the volume and pressure
can also be clearly observed. The similar results are shown in Fig.3
for CNT. The equation 3 now has been tested in many finite systems,
and in all cases, the constant pressure ensemble is guaranteed.

To show that the external pressure equals to the internal pressure (Eq.9 and 12), Fig.4 shows
the internal pressure as a function of external pressure for both $C_{60}$ and CNT.
For $C_{60}$ and CNT, the internal pressure is defined as Eq.9 and 12 respectively.
The simulation results clearly show that the constant pressure for both cases hold.
For other system, the similar results have been
obtained.

\begin{center}
{\bf $V$. The applications of ECPMD}
\end{center}

Materials under pressure have plenties of phenomena and attract
people for hundreds of years. The high pressure experiments provide
very important information relevant to the structure stability and
bonding of materials. Recently the studies on nano-systems under
pressure show fruitful new phenomena.\cite{Wickham00,Tolbert94,Tolbert95,Chen97,Jacobs01,Jacobs02,Zaziski04}
Promoted by the high pressure experimental work, Gong and his coworkers have
studied the structure and properties of nanosystems under pressures
by using ECPMD.  In the following of this review, the theoretical
approaches for the finite systems under pressure based on the ECPMD
are recalled, which includes,

(a)The elastic properties and melting behavior of metallic nanocrystals

(b)Structure transformation of CdSe nanocrystals

(c)Pressure induced  hard-soft transition of carbon nanotubes

\begin{center}
{\bf (a)The elastic properties and melting behavior of metallic nanocrystals}
\end{center}

 In the studies for metallic systems, the well-tested
many-body potentials are used, namely glue potentials for
Au,\cite{ercolessi88} Sutton-Chen potential for Ni,\cite{sutton} and
the tight-binding model for Ag.\cite{cleri} The
volume is calculated basing on \emph{VWS}, the system temperature is realized
by using Nos$\acute{e}$-Hoover thermostats.

The bulk modulus is one of the most important parameters of
materials, which reflects the the elastic properties of materials.
The elastic properties of Au, Ag and Ni nanocrystals have been
studied by using ECPMD.
Fig. 5 shows the pressure and energy as a function of reduced
volumes for Ni nanocrystals at 300K, and the counterpart of the bulk
phase calculated by APR-CPMD. In this figure, the energy is relative
to the minimum energies, and the volume is renormalized by
 the equilibrium volume at zero pressure and 300 K.
 Clearly nanocrystals is \emph{softer}
than the bulk phase, which is reflected by the larger volume change
for nanocrystal than bulk for the same applied pressure. The similar
results are also found for other metallic nanocrystals. The bulk
modulus can be obtained by fitting the energy-volume plot. The
obtained bulk modulus as a function of the size of nanocrystals is
shown in Fig.6, where the solid line is the linear fitting. It can
be seen that the MD data follows a straight line quite well,
which implys the elastic constants are reduced inversely with the size
of nanocrystal, similar to many other properties for nanosystems.
The bulk modulus for different temperature is also
obtained by fitting to the energy-volume plot. The obtained bulk
modulus as a function of temperature is shown in Fig. 7. The present
results show that, the bulk modulus decreases with the increase of
temperatures, which is similar to the bulk phase, and also consists
with the basic thermodynamics results.

The melting behavior is one of the common phenomena in nature, which
is also one of the most important process relevant to the properties
of materials. The basic thermodynamics shows that the melting
temperatures are strong affected by the external pressures, which is
characterize by the so-called Clapeyron equation for bulk materials.
Although the melting behavior of nanoclusters have been wildly
studied over the past decades, the pressure effect on the melting
behavior was not well understood for nano systems yet.
Recently Ye \emph{et al.} have carried out a detailed study for the
melting behavior of Ni nanoclusters under pressures.

Fig. 8 shows the melting points($T_{M}$) as a function of pressures
for $Ni_{561}$. Consistent with the basic thermodynamics, the
melting temperatures is increasing with the increasing of pressures.
The similar results is found for $Ni_{147}$ and bulk materials. The
latent heat versus pressure for $Ni_{561}$ cluster is shown in
Fig.9. From this figure, we can see that the latent heat seems to be
a constant in the studied range of pressures, where the average
value is about 0.0806eV. The volume difference between the solid
state and liquid state at the melting point versus pressure for
$Ni_{561}$ cluster shown in Fig.10. Assuming the latent heat is the constant,
The melting temperature and volume difference
are related through the Clapeyron equation quite well.

\begin{center}
{\bf (b) Structure transformation of CdSe nanocrystals}
\end{center}

Ye \emph{et al} studied the structure transformation of CdSe
nanocrystals using the ECPMD. The empirical potential developed by
Rabani.\cite{Rabani02} has been used to describe the interatomic
interaction. Most of their simulations are carried out at 300 K by
using a Nos\'{e}-Hoover thermostat.\cite{NH} The remarkable
structure character of nanoclusters is the large surface-volume ratio, thus it can
be expected that the surface could play an important role for the
structure transformation in nano systems.  In order to study the
effect of the surface structure on the transition mechanism, they
use nanocrystals of two different shapes, \emph{i.e,} the spherical
and faceted one, consisting of 500 to 5000 atoms. The initial
configuration of the spherical nanocrystal is simply cut from the
bulk CdSe of WZ structure. Faceted nanocrystals with well-defined
surface structure are obtained by cleaving the bulk lattice along
equivalent (100) WZ planes and at (001) and (00$\bar{1}$)planes
perpendicular to the [001] direction of the $c$ axis. The volume of
the nanocrystals is approximated based on finding the subset of atoms forming
the smallest convex polyhedron.

Ye \emph{et al} have observed the transformation from wurtzite to
rocksalt structure, but the process of transformation is strongly
dependent on the shape and size of the nanocrystals. Upon loading
the pressure, the spherical CdSe nanocrystals is found to directly
transfer to rocksalt structures with nanoscale grain boundary
formed, while the faceted ones can first transfer to hexagonal MgO
structure, and then the final rocksalt structure with grain boundary
free. These results are similar to that calculated by the direct method for the same systems.

Fig.11 shows their calculated the volume-pressure plot. From this
figure, it clearly indicates the structural transformation of the
CdSe nanocrystal up loading pressures. The volume of the spherical
$Cd_{502}Se_{502}$ nanocrystal decreases smoothly with increasing
pressure up to a critical pressure$\sim$ 8.0 GPa, at which the
volume decreases abruptly as a result of the transformation from WZ
to RS(left penal of Fig.14). This is in good agreement with the high
pressure experiment for the same system.\cite{Chen97} For faceted
nanocrystals, an intermediate
structure($\textbf{C}$$\rightarrow$$\textbf{D}$) clearly exists
between WZ and RS. Detailed analysis of the variations of
coordinations shows that the WZ structure of faceted nanocrystal
transforms to a five-fold coordinated structure around 1.4GPa. The
five-fold coordinated structure has been reported as a stable phase
of MgO under hydrostatic tensile
loading.\cite{Limpijumnong01,Kulkarni06} When the pressure continues
to increase, the five-fold coordinated structure transforms to
six-fold coordinated RS structure. The transition process is also
found to be highly hysteretic. Upon pressure releasing, The rock
salt structure remains stable down to pressures significantly below
the observed "upstroke" transition pressure. As low as 0.5 GPa, the
sample begin to restore WZ structure. At atmospheric pressure, the
WZ is recovered but with a few defects near surfaces.

For spherical nanocrystals, the transformation pressure decreases
with size increasing. This trend coincides with the experimental
results of Alivisatos.\cite{Tolbert94,Tolbert95} (Note, in Ye
\emph{et al}'s studies,  "upstroke" transition pressure is used as
the transition pressure) In contrast to the spherical one, the
transformation pressures (both from four-fold to five-fold and from
five-fold to six-fold one) of facet nanocrystals increase with the
increasing of crystal sizes. Fig. 12 and Fig.13 shows the
transformation pressure as a function of sizes for facet
nanocrystals.  The different dependence of transformation pressure
on the size has been discussed based on thermodynamics
considerations by including the surface effect.

For all spherical nanocrystals, nano-scale 'grain' boundaries are
formed during and after the transformation(see Fig.14). As the size
of the nanocrystal increases, the multiple grains phenomena become
more and more obvious. In contrast, the facet one is almost grain
boundary free. The generation of grain boundary has been discussed
based on the nucleation mechanism.\cite{Ye_un}

\begin{center}
{\bf (c) Pressure induced  hard-soft transition of carbon nanotubes}
\end{center}

Gong and his co-workers have presented a detailed investigation on
the behavior of carbon nanotubes under hydrostatic pressures by
ECPMD.\cite{Sun04,Ye05,Ye07a} In their simulation, a few
single-walled carbon nanotube(SWCNT), double-walled carbon
nanotubes(DWCNT) and multi-walled carbon nanotube(MWCNT) are
investigated. For DWCNT, both commensurate and incommensurate one
are considered. DWCNTs consisting of tubes, which has the same
chirality, are typically commensurate; otherwise they are
incommensurate. The volume is calculated by the minimal polyhedron method.
The periodic boundary condition
in the axial direction, and free boundary condition in the radial
directions are used. The interaction between carbon atoms is
described by a parameterized many-body potential.\cite {Te,Br} The
intertube and intratube van der Waals interaction are modelled by
the Lennard-Jones (LJ) potential.\cite{Girifalco56,Henrard99} To
confirm the results of the classical molecular dynamics method, they
also have repeated some calculations by {\it ab-initio} molecular
dynamics method.

They found that all studied nanotubes(SWCNTs, DWCNTs and MWCNTs) undergo a
pressure-induced hard-to-soft phase transition. The $hard$ phase at
low pressure exhibits a typical bulk modulus of $~100$ GPa, while
the $soft$ phase at high pressure exhibits a bulk modulus of only
$\sim$1 GPa. Fig. 15 shows the pressure and the total energy as a
function of reduced volume for a (10,10) nanotube at 300 K, where
the energy at zero pressure is set to zero. Clearly, a transition at
$\sim$1.0 GPa is observed. Below the transition, the $hard$ phase
has a radial compressibility of 0.01 GPa$^{-1}$. Above the
transition, the $soft$ phase has a radial compressibility about two
orders of magnitude larger. The similar behavior was observed for
other tubes.

After the hard-to-soft  transition, the cross section of nanotubes changes
from circular to elliptical shape. The evolutions of the
cross-section shape, the bond length, and the bond angle with
increasing pressure for a (10,10) nanotube are shown in Fig. 16.
where two
principal axes (long axis $a$ and short axis $b$) are used to
characterize cross section. Below the transition pressure, $a$
remains almost equal to $b$, defining a circular shape( see
Fig.16). Above the transition pressure, $a$ becomes larger than
$b$, defining an elliptical shape. Eventually, as the long axis
continues to increase and the short axis continues to decrease, the
elliptical shape undergoes another transition to a dumbbell shape.
Under even higher pressure, the dumbbell tube can become so flat
that the spacing between the opposite side walls approaches the
layer spacing in the graphite ($\sim$3.35$\AA$).

The trend of change in bond length and bond angle provides a good
explanation of the hard-to-soft transition. Fig. 16(b) shows that
the percentage change in bond length and bond angle increases
simultaneously with increasing pressure below the transition,
indicating a uniform shrinking of the circular shape under pressure.
Above the transition, the bond length remains unchanged but the
change of bond angle increases sharply with increasing pressure.
Since it costs much more energy to change bond length than to change
bond angle. Below the transition, the structural response to the
external pressure is largely taken by the changing bond length of a
circular shape, giving rise to a hard phase; while above the
transition, the structural response to the external pressure is
largely taken by the changing bond angle of an elliptical shape,
giving rise to a soft phase.

The critical transition pressure depends strongly on the tube
radius. Fig. 17 shows the simulated transition pressures as a
function of tube radius for SWNTs(solid dots). The smaller the
radius, the higher the transition pressure. To understand the above
simulation results, they also provide a general analysis based on
continuum elastic theory. According to their deduction, transition
pressure $P_{t}$ has,

\begin{equation}
P_t\approx\frac{3D}{R_0^3},
\end{equation}
where \emph{D} is the constant related to the elastic properties of
NTs, and $R_{0}$ is the tube radius. This analytical dependence of
$P_t$ on $R_0$ is in very good agreement with the MD simulations, as
shown in Fig. 17.

The bulk modulus of the {\it hard} phase follows,

\begin{equation}
B_h=\frac{C}{2R_0}.
\end{equation}
where \emph{C} is another constant related to the elastic properties
of NTs. This analytical dependence of bulk modulus on $R_0$ are in
very good agreement with the MD simulations, as shown in Fig. 17.

The similar pressure-induced structural transition has been found
for all studied DWNT's. Comparing with SWNT, the transition pressure
of DWNT is much enhanced, but it still follows the Eq. 20. Figure 18
presents the transition pressure as a function of the both inner and
outer tube radius. The results imply that the van der Waals forces
between two tubes does affect the transition pressure. The
transition pressure of a outer tube in DWNT can be increased largely
by inserting an inner tube. In fact, the DWNT can be considered as a
psudo-single-walled nanotube with effective thickness, which should
be larger than the real SWNT.

The remarkable feature of MWNTs is its encapsulation effects,
especially, when the system is undergoing pressure, the outer shell
acts as protector for the inner shell. They found that, the response
pressure of inner tube is much smaller than the external pressure, while
the response pressure of outer tube is much
closer to the external pressure. To characterize the pressure
transmission, Ye \emph{et al} define a response pressure for the
tubes.\cite{Ye07a} Left panel of Fig. 18 shows the
response pressure as functions of external pressure of (5,5)@(10,10)
DWCNT. Form this figure, on can see that the response pressure of
both inner and outer tube increases linearly with the external
pressure below 8 GPa, while the value for inner tube is about two
times smaller than outer one. Interestingly, for all the studied
tubes, when the external pressure is higher than a certain
value, at which the hard-to-soft transition happens, ($\sim$8 GPa for (5,5)@(10,10) and $\sim$7 GPa for
(5,5)@(10,10)@(15,15)), the response pressure of inner tube increases
sharply, while increasing of the response pressure of the outer tube
slows down.

Ye \emph{et al} assume a linear relationship between the response
pressure and the external pressure before the structural transition
happens, they define pressure transmission efficiency $\beta$ by
\begin{displaymath}
P_{r}=\alpha+\beta P_{e}
\end{displaymath}
where $P_{r}$ and $P_{e}$ are the response pressure of the inner
tube and the external pressure respectively, $\alpha$ is the
response pressure of the inner tube without external pressure. Fig.
19 presents the pressure transmission efficiency of (n,n)@(n+5,n+5)
DWCNTs as a function of the radius of the outer tube. The pressure
transmission efficiency is found to increase with the tube radius.

In contrast to commensurate DWCNTs, the incommensurate DWCNTs have
lower transmission efficiency. The pressure transmission efficiency
are 0.30 and 0.35 for (6,6)@(19,0) and (10,10)@(26,0) respectively,
while the pressure transmission efficiency of their commensurate
counterpart (6,6)@(11,11) and (10,10)@(15,15) are 0.35 and 0.43
respectively. Obviously the morphology combination does affect the
vdW interaction between inner and outer tubes. The calculations show
that the pressure transmission of the commensurate DWCNTs is more
efficient. This might be due to the fact that the atomic positions
in adjacent shells are well matched in commensurate DWCNTs,
meanwhile the intralayer vdW force favors commensurate
tubes.\cite{Kolmogorov00}

\begin{center}
{\bf $VI$. Summary}
\end{center}

By writing the volume of a system as a function of coordination of
atoms and extending the Lagrangian of the system to include a PV
term, a constant-pressure
molecular-dynamics method can be achieved in a simple but physically rigid way.
This method is different the traditional constant-pressure one by treating volume
as a part of potential in steady of a generalized dynamics variable.
This method is specially suitable for finite systems and the system without periodic boundary conditions.
 In this paper, the
varies of application and some technique key issues of this method
are reviewed. The method is fairly general and can find widespread
applications.

\begin{acknowledgments}
This research is partially supported by the National Science
Foundation of China, the special funds for major state basic
research and Shanghai Project for the Basic Research. D.Y.S is also
partially supported by Shanghai Municipal Education Commission and
Shanghai Education Development Foundation, and the Pujiang Project
of Shanghai Municipal. The computation is performed in the
Supercomputer Center of Shanghai, the Supercomputer Center of Fudan
University and CCS.
\end{acknowledgments}

\newpage
\begin{figure}[fig1]
\centering
\includegraphics[width=100mm]{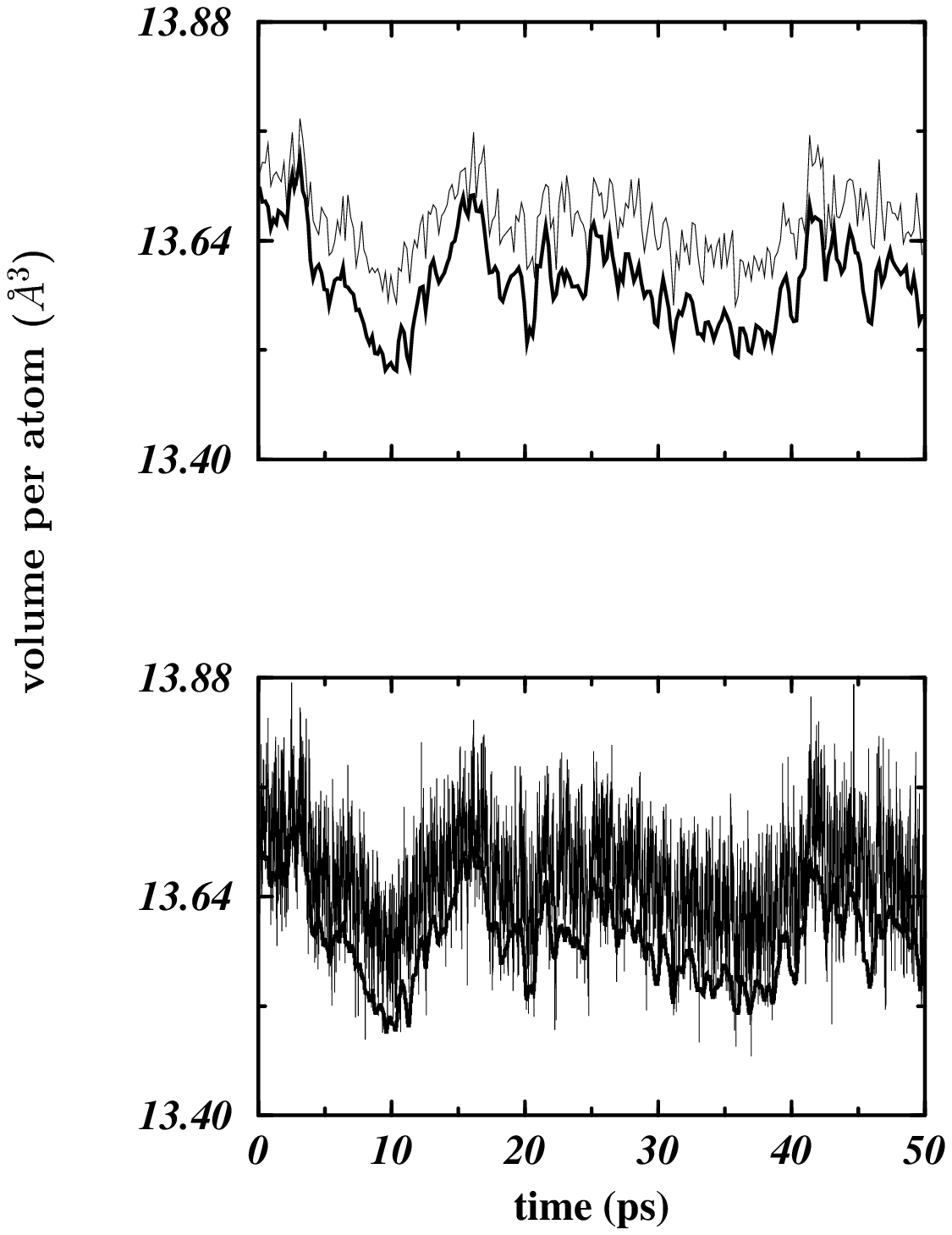}
\caption{Bottom panel: The volume per atoms for liquid Ni calculated
by VWS (thin line) and the exact one (thick line) obtained by
traditional constant pressure molecular dynamics simulation at 3000K
and 5GPa. Up panel: the same as bottom panel except the volume
averaged over a short time. The two lines
are close to each other }
\end{figure}

\begin{figure}[fig2]
\centering
\includegraphics[width=100mm]{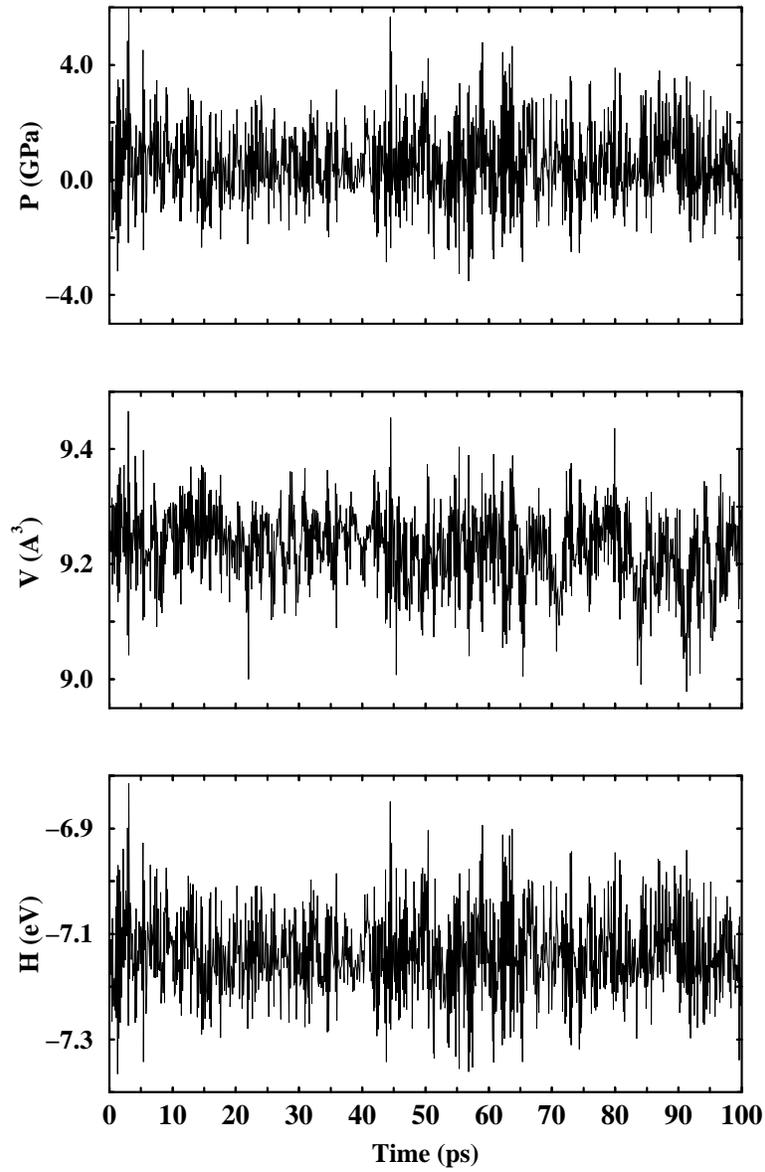}
\caption{The evolution of the instantaneous internal pressure
(middle penal), enthalpy (low penal) and volume (up penal) during ECPMD
runs for $C_{60}$. The ECPMD does recover a constant pressure
simulation. }
\end{figure}

\begin{figure}[fig3]
\centering
\includegraphics[width=100mm]{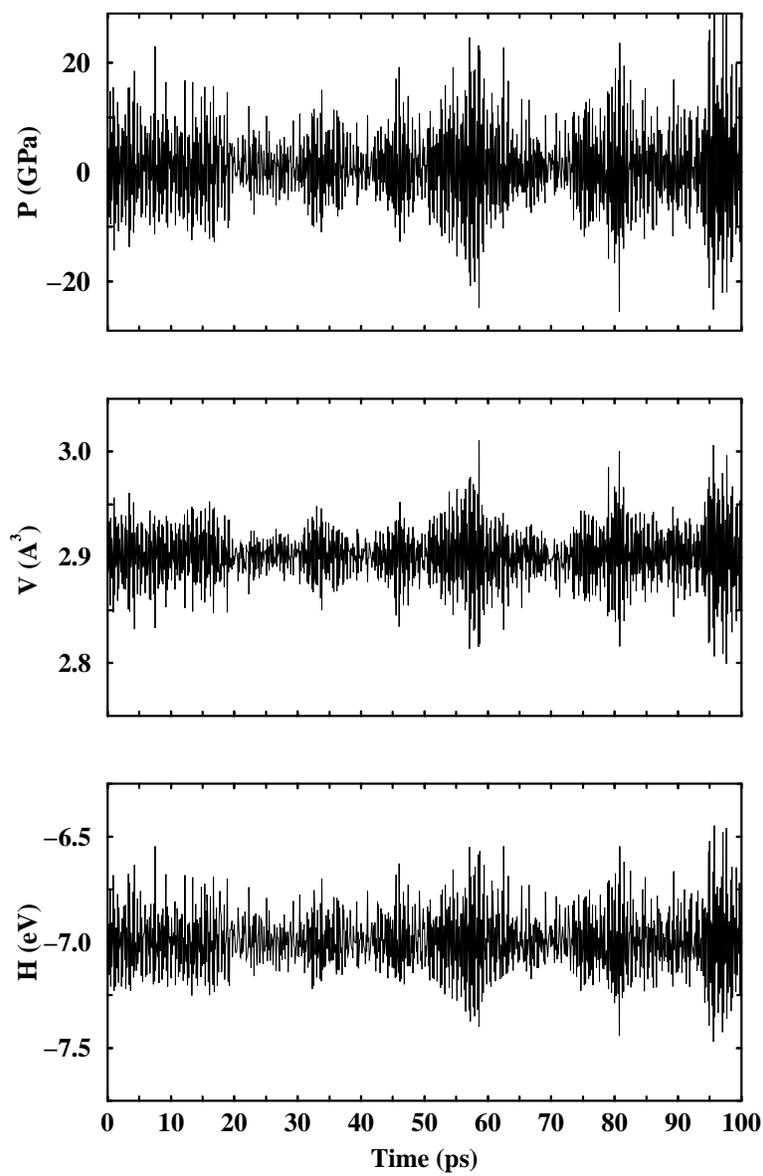}
\caption{The same as Fig.2 except the system is carbon nanotube.}
\end{figure}

\begin{figure}[fig4]
\centering
\includegraphics[width=100mm]{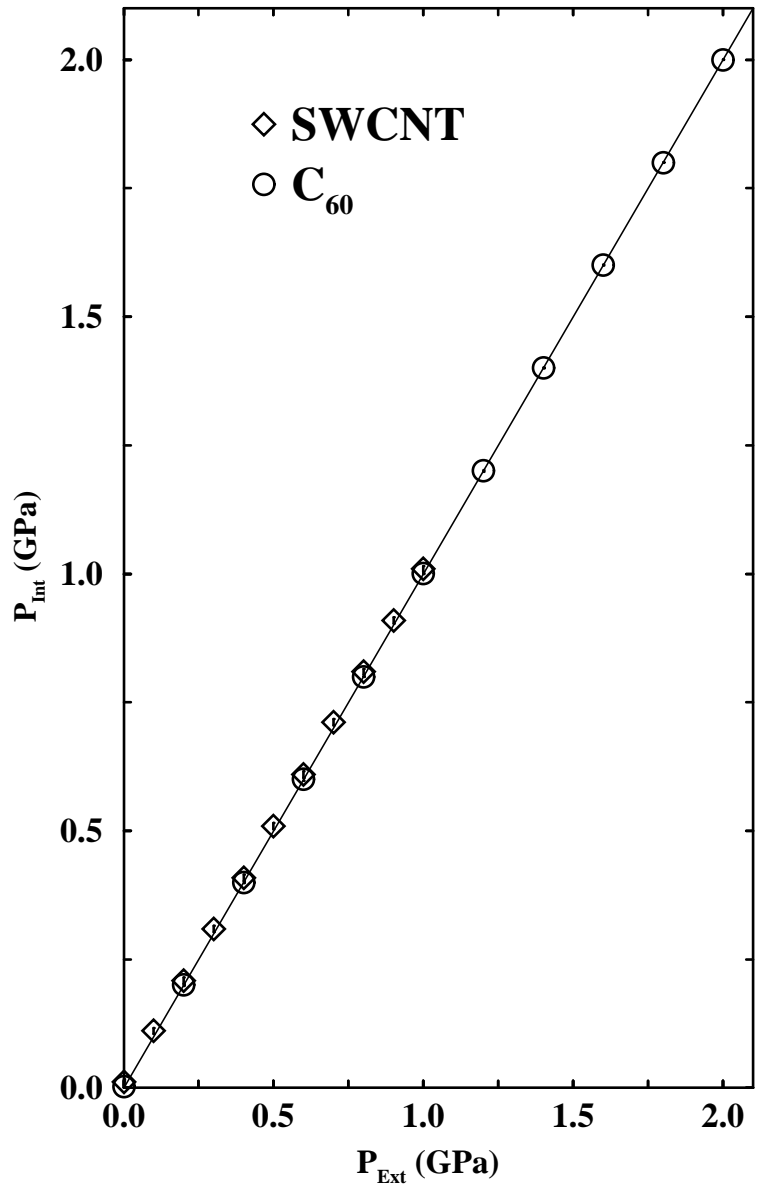}
\caption{ The internal pressure as a function of external pressure for $C_{60}$ and carbon nanotube.
Clearly they are equal!}
\end{figure}

\begin{figure}[fig5]
\centering
\includegraphics[width=100mm]{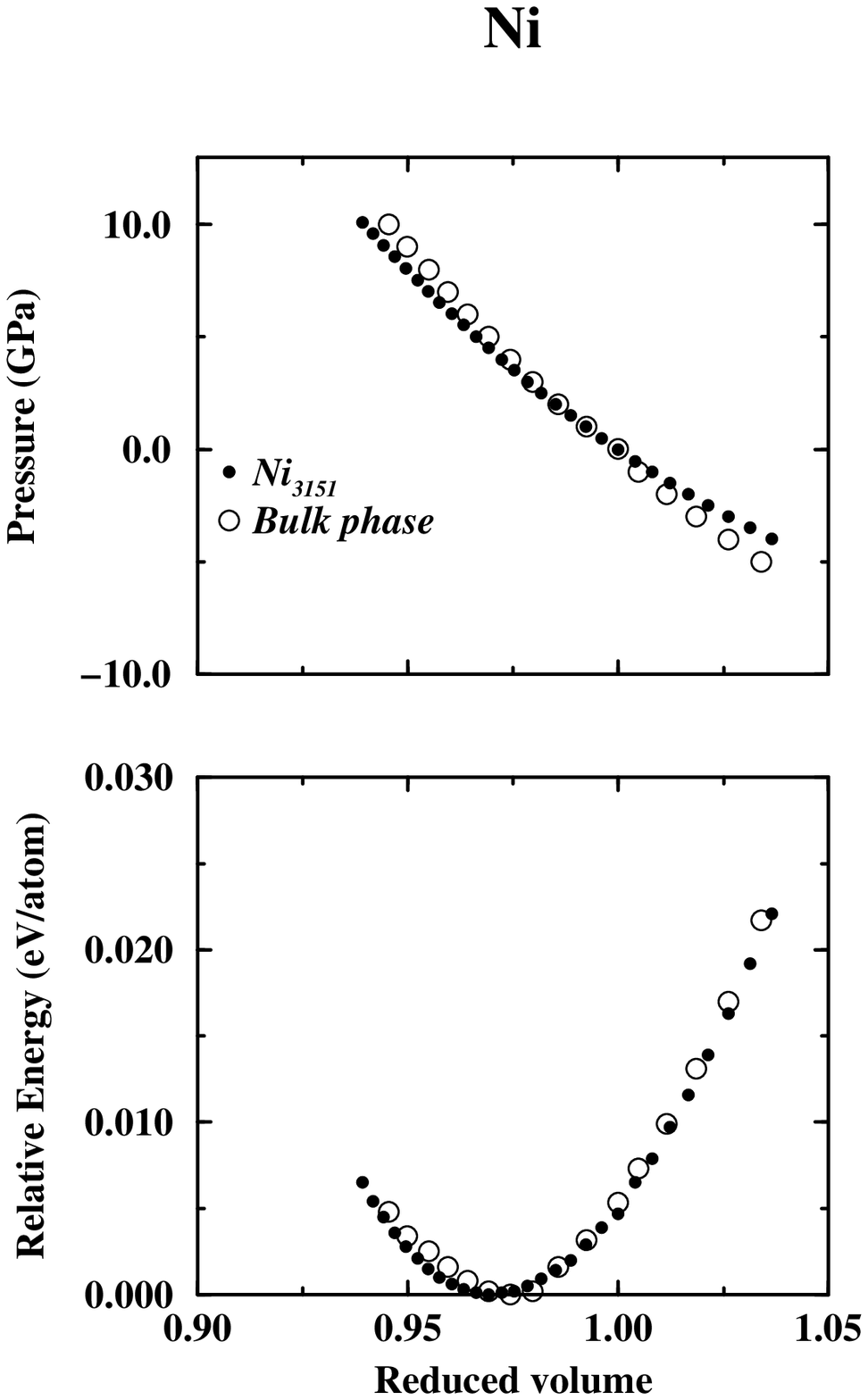}
\caption{ The equations of state for Ni nanocrystals (filled
circles) and bulk phase (open circles), where the data for
$Ni_{3151}$ nanocrystals and bulk phase are calculated by ECPMD
traditional molecular dynamics simulation respectively. }
\end{figure}

\begin{figure}[fig6]
\centering
\includegraphics[width=100mm]{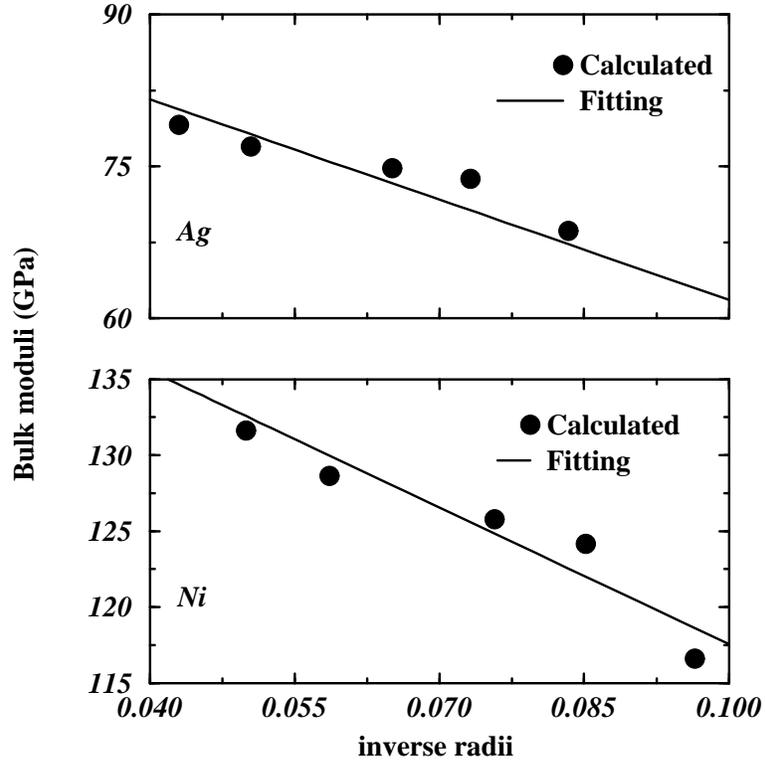}
\caption{ The bulk modulus as a function of the size of
nanocrystals. Circles: calculated data, line: the linear fitting to
the data. }
\end{figure}

\begin{figure}[fig7]
\centering
\includegraphics[width=100mm]{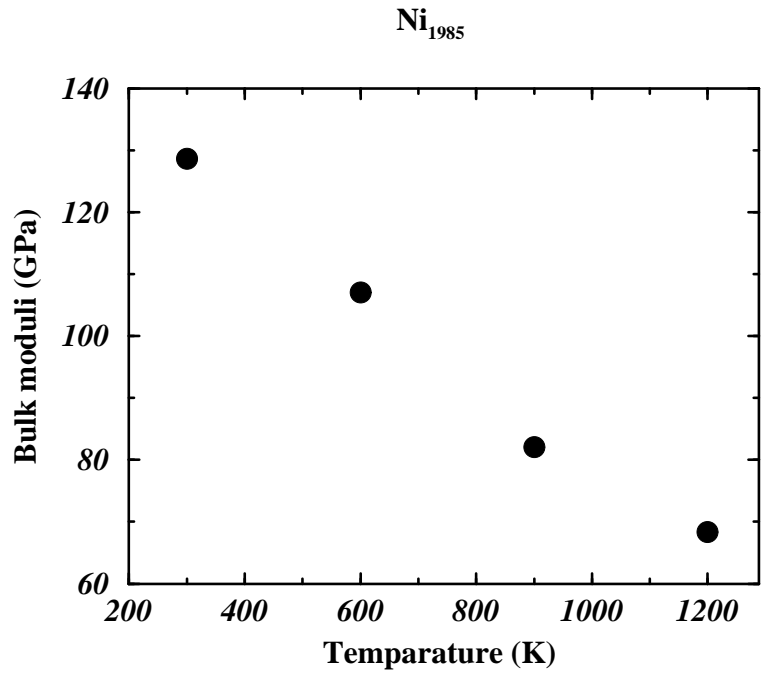}
\caption{ The bulk modulus as a function of temperature for
$Ni_{1985}$. }
\end{figure}

\begin{figure}[fig8]
\centering
\includegraphics[angle=-90,width=100mm]{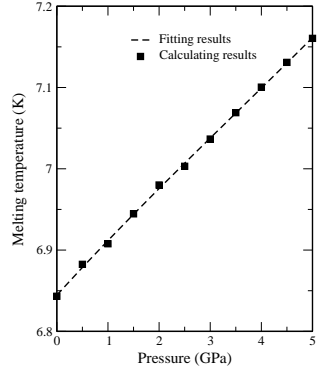}
\caption{ The melting points($T_{M}$) as a function of pressures for
$Ni_{567}$. The melting temperatures is found to increase with the
increasing of pressures, in agreement with the basic
thermodynamics.}
\end{figure}

\begin{figure}[fig9]
\centering
\includegraphics[angle=-90,width=100mm]{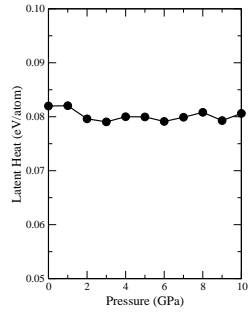}
\caption{ The latent heat versus pressure for $Ni_{561}$ cluster.
The latent heat seems to be a constant in the studied range of
pressures. }
\end{figure}

\begin{figure}[fig10]
\centering
\includegraphics[angle=-90,width=100mm]{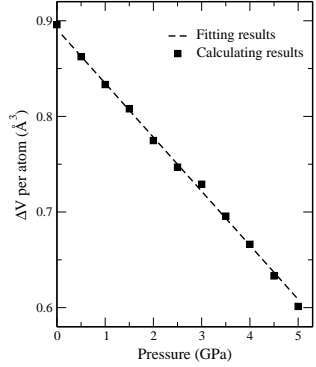}
\caption{The volume difference between the solid state and liquid
state at the melting point versus pressure for $Ni_{561}$ cluster.
Dash line: deduced from Clapeyron equation, Square: from the
simulation. }
\end{figure}

\begin{figure}[fig11]
\includegraphics[width=100mm,angle=270]{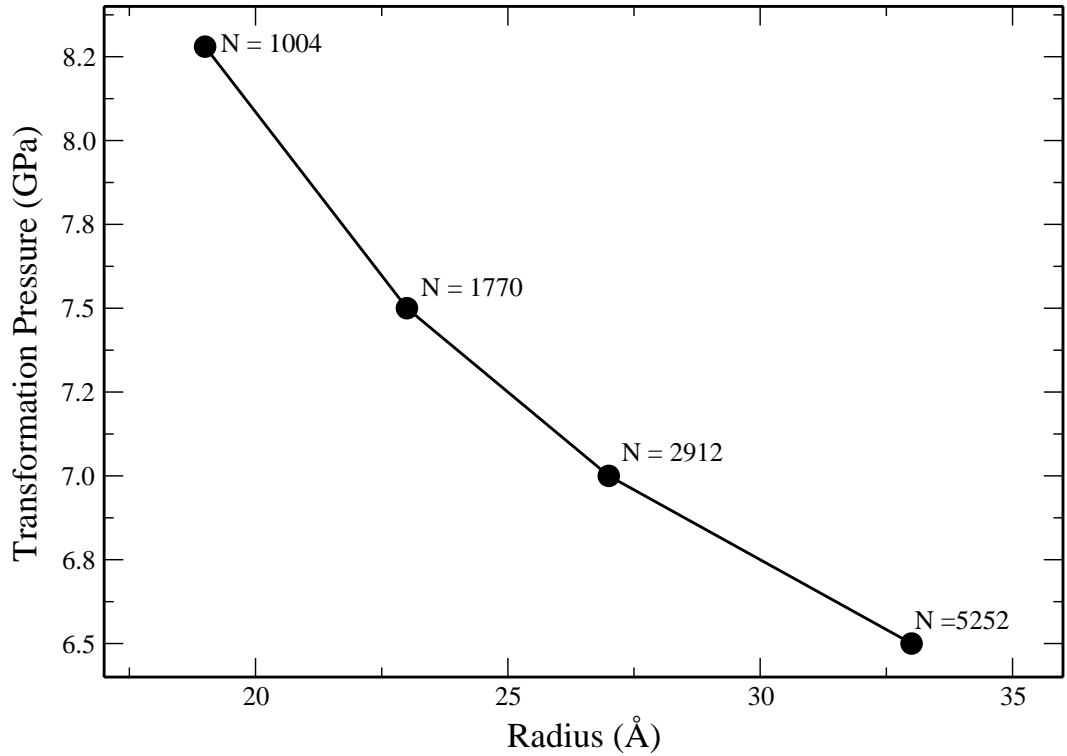}
\caption{ Volume versus pressure for faceted $Cd_{1162}Se_{1162}$
(right) and spherical $Cd_{502}Se_{502}$ nanocrystal (left). The
discontinuity of the slopes indicates structural transformation.}
\end{figure}

\begin{figure}[fig12]
\includegraphics[width=100mm,angle=270]{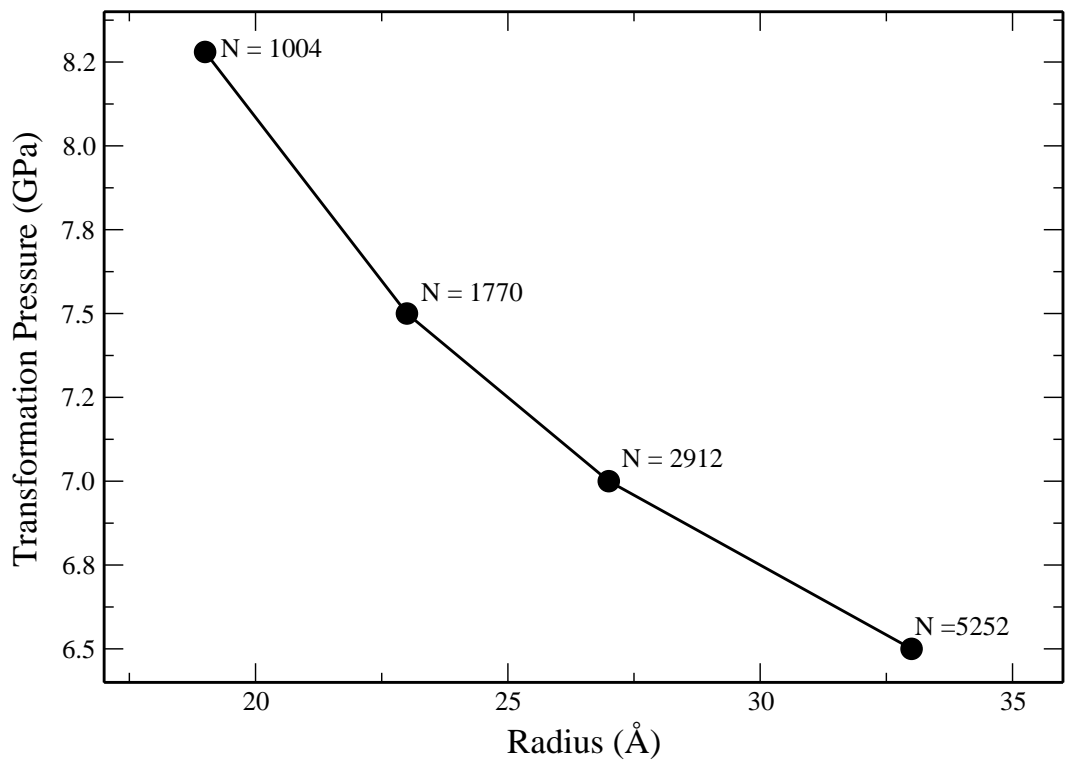}
\caption{\label{fig:epsart}Variation of the transformation pressure
with radius for spherical CdSe nanocrystals at 300 K. With increasing nanocrystal
size, the transformation pressure decreases.}
\end{figure}
\newpage

\begin{figure}[fig13]
\includegraphics[width=100mm]{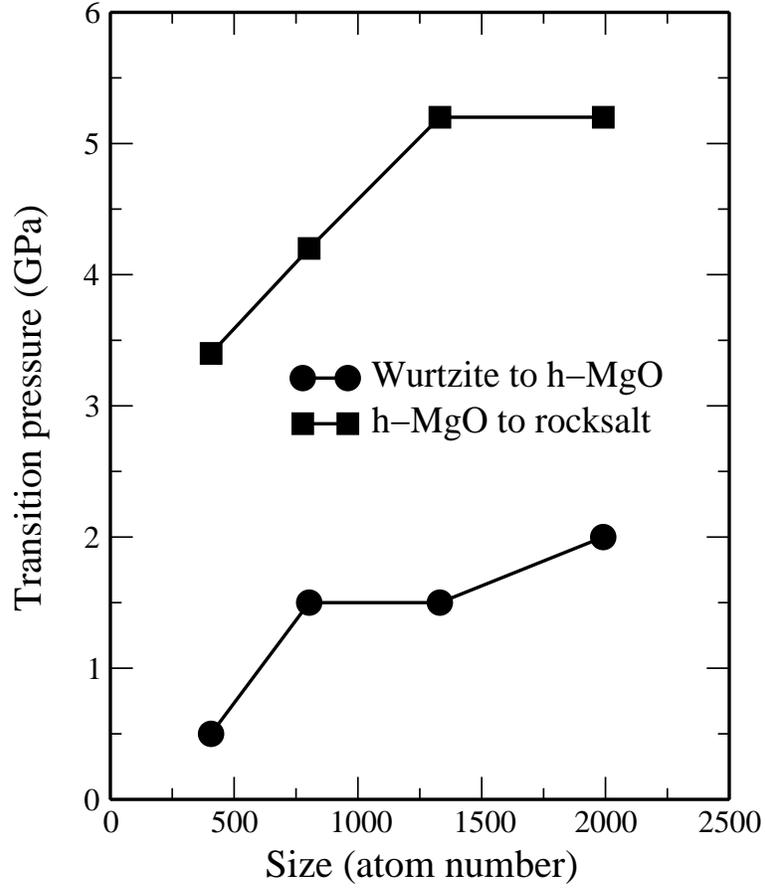}
\caption{\label{fig:epsart}The transformation pressure
as a function of radius for facted nanocrystals at 300 K. In contract with
spherical one, the transformation pressure increases with increasing
nanocrystal size.}
\end{figure}

\begin{figure}[fig14]
\includegraphics[width=70mm]{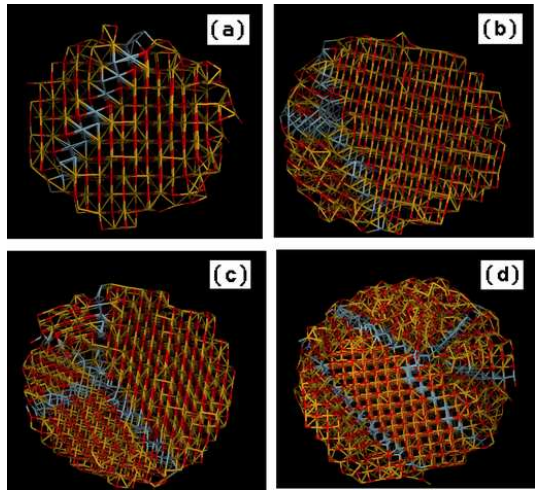}
\caption{\label{fig:epsart}Domains after the structural
transformation in spherical nanocrystals. (a), (b), (c) and (d) show
the cross sections through the middle of the nanocrystals of radius
19 {\AA}, 23 {\AA}, 27 {\AA} and 33 {\AA}, respectively. The grain
boundaries are shown with gray atoms.}
\end{figure}

\begin{figure}[fig15]
\includegraphics[width=100mm]{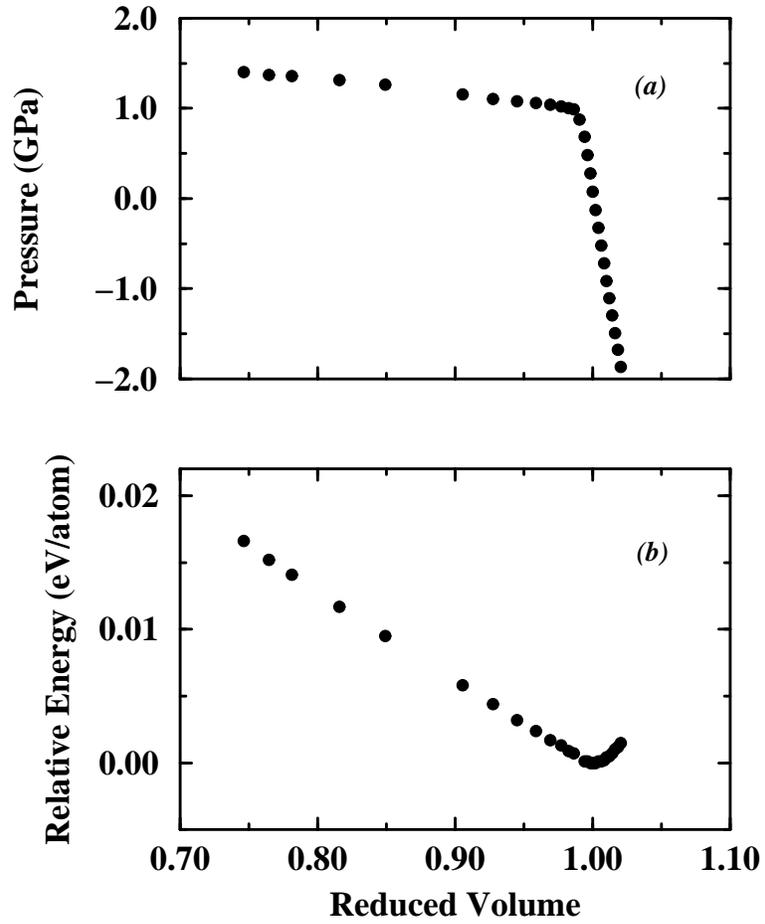}
\caption{The energy and pressure as a function of the reduced volume
for (10,10) carbon nanotube at 300 K. The minimum energy is set to zero,
and the volume is normalized by the equilibrium
volume without the external pressure. At about 1.0 GPa, the $hard$
phase with bulk modulus of about 100 GPa transforms into the $soft$
phase with bulk modulus of just a few GPa.}
\end{figure}

\begin{figure}[fig16]
\includegraphics[width=100mm]{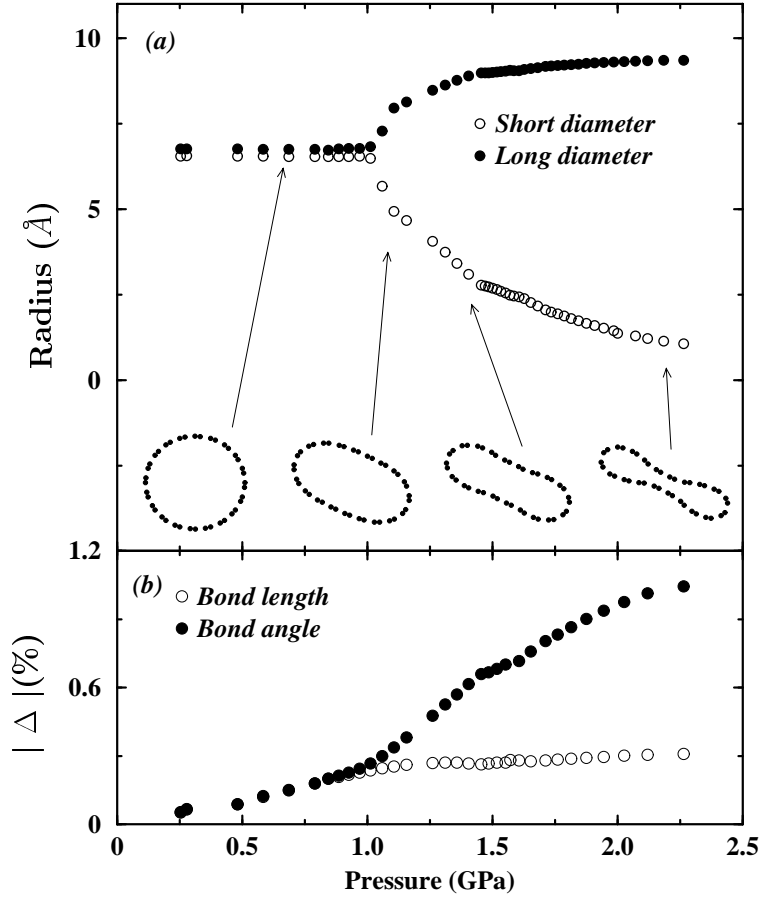}
\caption{The length of the long and short axes, as a function of
pressure for (10,10) nanotube. The shape of cross section at some
selected pressures is plotted at the bottom of the figure. The
absolute relative change of bond length and bond angle as a function
of pressure for (10,10) nanotube, the data is obtained by quenching
the system from 300 K to 0 K at constant pressure.}
\end{figure}

\begin{figure}[fig17]
\includegraphics[width=100mm]{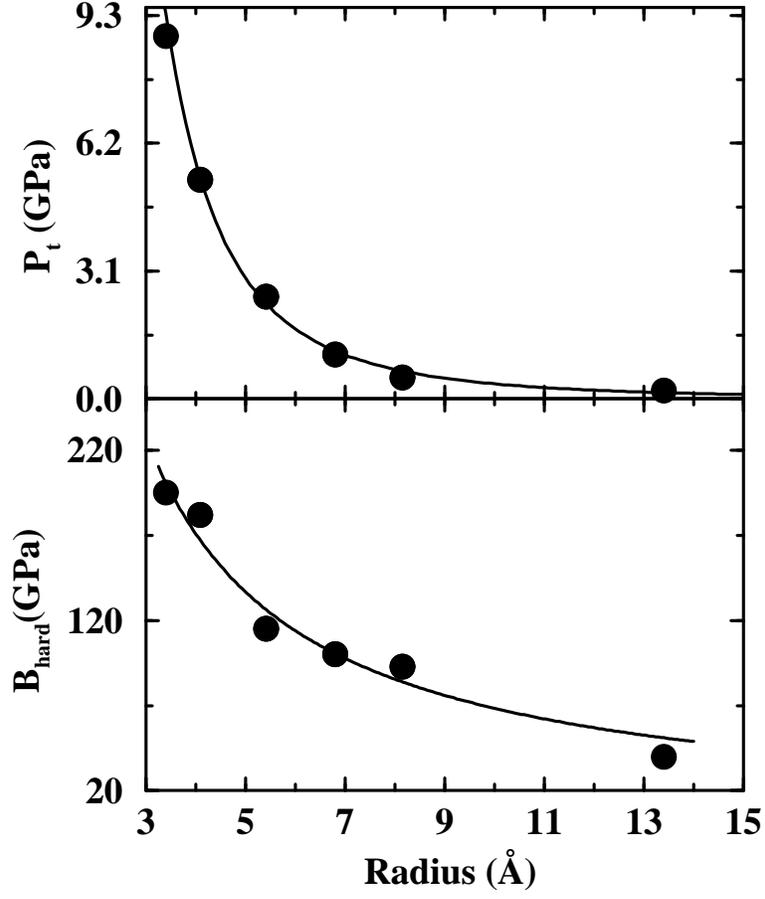}
\caption{The transition pressure (upper panel) and the elastic
modulus (lower panel) as a function of tube radius at 300 K. The
solid line is a least-square fit to the data using Eq. 20 (upper
panel) and Eq. 21 (lower panel). The simulated data follows nicely
with the predicted behavior.}
\end{figure}

\begin{figure}[fig18]
\includegraphics[width=100mm]{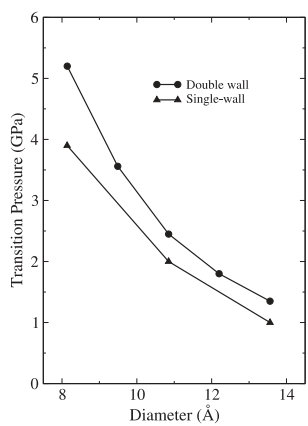}
\caption{Transition pressure as a function of radius of nanotubes at
300 K for a few DWCNTs. }
\end{figure}

\begin{figure}[fig19]
\includegraphics[width=100mm]{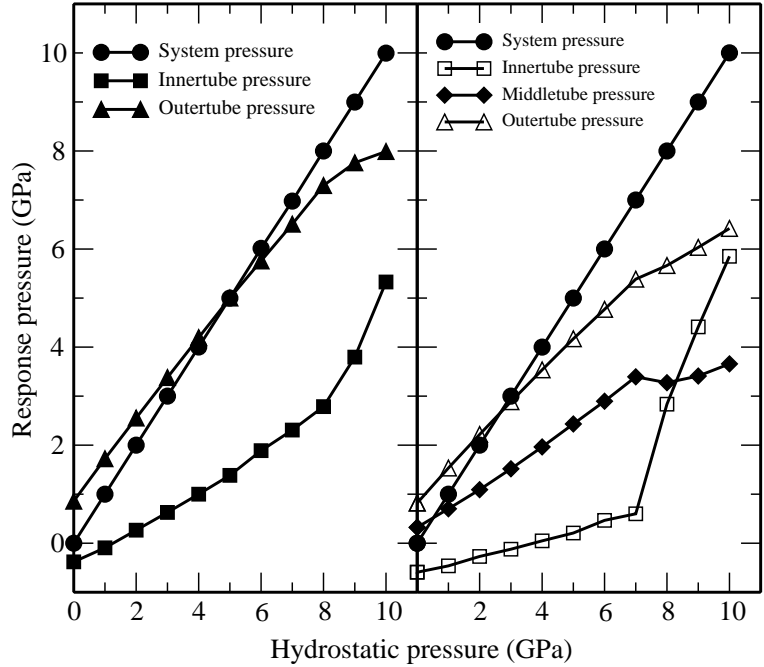}
\caption{Response pressure as a function of the external pressure
for (5,5)@(10,10) DWCNT(left) and (5,5)@(10,10)@(15,15)
TWCNT(right). The system response pressure is exactly the same as
the external pressure. The response pressure of inner tube is much
smaller than that of the outer one before structural transition
occurs.}
\end{figure}

\begin{figure}[fig20]
\includegraphics[width=100mm]{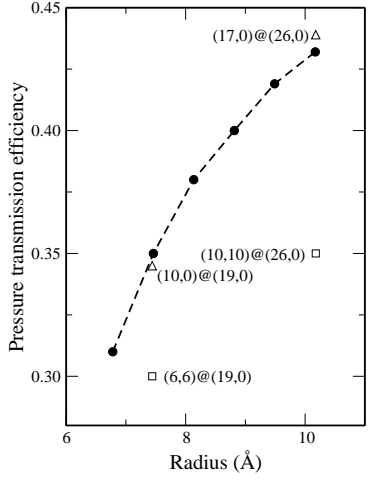}
\caption{Pressure transmission efficiency of commensurate
(n,n)@(n+5,n+5) DWCNT (with n=5,6,7,8,9,10) versus the outer tube
radius. The transmission efficiency increases with the tube radius.
The incommensurate DWCNTs (6,6)@(19,0) and (10,10)@(26,0) is also
shown for comparison, of which the transmission efficiency is
smaller.}
\end{figure}
\end{document}